# The NOT GL survey for multiply imaged quasars

**A.O. Jaunsen**[1]*, **M. Jablonski**[2], **B.R. Pettersen**[3], and **R. Stabell**[1]

[1] Institute of Theoretical Astrophysics, University of Oslo, P.O.Box 1029, Blindern, N-0315 Oslo, Norway.
[2] Nordic Optical Telescope, Observatorio Roque de los Muchachos, Apartado 474, E-38700 Santa Cruz de La Palma, Spain.
[3] Statens Kartverk, N-3500 Hønefoss, Norway

January 31, 1995

**Abstract.** A gravitational lens (GL)-search program, initiated in 1990 at the Nordic Optical Telescope (NOT), has revealed several possible GL-candidates among a sample of 168 quasars (QSOs), chosen from three lists compiled by C. Hazard, D. Reimers and J. Surdej, respectively. Some of these candidates, selected for having close companions (within 5 arcseconds), were imaged in several filters and their colours compared. Low dispersion spectra of the most promising candidates were also obtained at the NOT and ESO New Technology Telescope (NTT). None of these has proved to be strong candidates of gravitational lensing effects. We present this new sample of QSOs and combine it with previously published optical QSO samples in a statistical analysis to yield constraints on flat cosmologies and galaxy velocity dispersions. Finally, by simulating larger samples of quasars and gravitational lenses, we discuss how the uncertainties affecting our present results would be changed.

## 1. Introduction

One of the (several) aims of GL-surveys is to perform statistical analysis of the lensing frequency and of the lensed image configurations (morphology), to derive various fundamental parameters for a universe model. In Section 2, we discuss and present our sample of QSOs. We first discuss the correction for GL incompleteness in the QSO samples and define the principles leading to our selection of GL-candidates. Our chances of finding GL-systems depend on how the sample of QSOs is chosen. Furthermore, the lensing probability is dependent upon the detectability limits imposed by the instruments and observational conditions at the site. In Section 3, we investigate our abilities to detect GL-systems and characterise the NOT angular selection function, defining the area (in the magnitude difference-angular separation plane) within which a pair of point-like images is considered to be a possible GL candidate and used in the analysis. In Section 4, we introduce the Singular Isothermal Sphere (SIS) galaxy model, employed in our statistical model. Furthermore, we also include the important magnification bias, which accounts for the increase in the observed frequency of lensing. The possible GL-candidates found in the NOT sample are presented in Section 5, where we also give a brief discussion of each candidate. Finally in Section 6 we present the total combined sample and the constraints our computations impose on galaxy velocity dispersions and on the baryonic-matter density parameter $\Omega_M$ (see Carroll et al. 1992). We also briefly comment on the influence of dark matter (DM) on velocity dispersions in gravitational lensing statistics. The decrease in the uncertainties of our results when the sample is artificially increased are investigated by applying the statistical model on synthetic samples.

## 2. The QSO sample

The observed QSO sample (listed in Table 1 and illustrated in Fig. 1) was selected from 3 different QSO lists provided by C. Hazard, D. Reimers and J. Surdej, respectively. The latter one consists of QSOs chosen from the Véron-Cetty & Véron (1991) catalogue, while the other two result from independent QSO surveys. Some of the QSOs in the two first samples were difficult to identify in the images taken, due to crowded CCD-fields. To securely identify the QSO in such cases, we have obtained finding charts from the digitised Palomar Atlas with a magnitude cutoff at about $V \simeq 21$. This allowed us to compare the observed field with that of the Palomar Atlas and determine the presence (or non-presence) of the target QSO.

Most of the larger optical GL-surveys optimise the effectiveness of finding GLs in their surveys by including intrinsically bright QSOs ($M_V \leq -27$). Such QSOs are often called highly luminous quasars (HLQs) (Surdej et al. 1987). The strong correlation between lensing probability and abso-lute magnitude is discussed in Surdej et al. (1993). Most of the QSOs from our sample (marked by ∗ in Fig. 1) tend to lie beneath the line indicating absolute magnitude $M_V = -27$. The irregular pattern of the absolute magnitude lines is due to the correction for emission lines appearing in the V-filter for various redshifts. The emission line correction, K-correction and computation of absolute magnitudes are discussed in Véron-Cetty & Véron (1991).

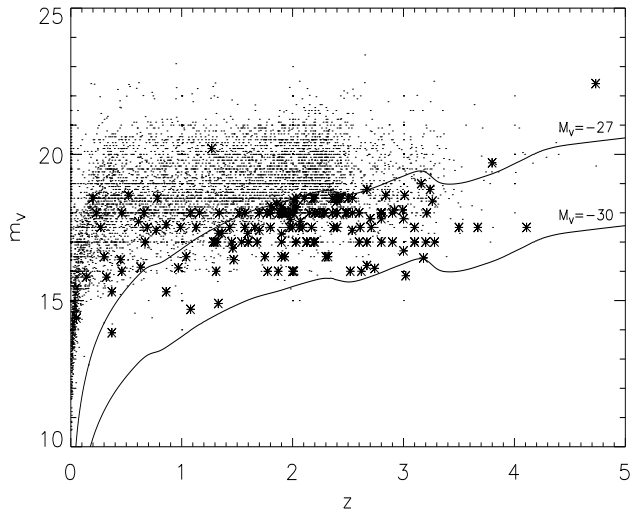

**Fig. 1.** Magnitude-redshift distribution of the 168 QSO NOT sample, each object indicated by ∗. The small dots represent the QSOs from the Véron-Cetty & Véron (1991) and Hewitt & Burbidge (1993) catalogues. The solid lines indicate the absolute magnitudes $-27$ & $-30$ for an Einstein-de Sitter universe model ($\Omega_M = 1, \Omega_\Lambda = 0$).

The selection of a flux limited sample of HLQs to maximise the chances of finding GLs is consistent with the previously presented samples by Surdej et al. (1993), Crampton et al. (1992), Yee et al. (1993) and Maoz et al. (1993) (hereafter the SCYM sample). Among the 645 HLQs from this sample, only 31 were duplicated in the 168 HLQ NOT sample, leaving 137 unduplicated quasars. The median apparent magnitude of the NOT sample is $m_V \sim 17.5$, the median redshift is $z \sim 2.0$ and the median absolute magnitude is $M_V \sim -27.7$ (using $\Omega_M = 1, \Omega_\Lambda = 0$ and $H_0 = 50$ km s$^{-1}$ Mpc$^{-1}$).

Among the lens surveys conducted in the previous years, most of the objects have been selected from already existing QSO catalogues. It has been suggested that many QSO catalogues are biased against the inclusion of elongated or diffuse QSOs, because of the selection routines used in search for QSOs. According to Kochanek (1991), some GL candidates and real GL-systems are lost in surveys searching for GLs as a direct consequence of the inherent selection effects in QSO catalogues. Elongated QSOs are interesting candidates to gravitational lensing, because closely separated GL-systems would appear elongated under (sub-)critical seeing conditions [1].

Depending on the seeing conditions, which dictate the resolution and detection of small separation GLs at ground-based observatories, this latter selection effect might play an important role in GL-surveys. We will therefore conduct tests through simulations in Section 3 to estimate the effects of limited resolution on the detectability of candidates and determine an angular selection function which describes the validity of our observational conclusions.

Furthermore, a colour criterion used in some QSO surveys may bias the selection of QSOs. A QSO superposed on a (lensing) galaxy is subject to a reddening effect (from the red colour of the galaxy) which may endanger a lensed QSO from being recognised as a QSO, thereby affecting the completeness of the QSO catalogue. This effect was pointed out by Kochanek (1991), where as much as 30% of the gravitationally lensed QSOs are believed to be excluded from QSO catalogues due to such effects. Surdej et al. (1993) corrected for this effect in their computations of the lensing frequencies from GL surveys, thus multiplying the optical depth by a correction factor $S_{cat} = 0.7$. This factor is included in the present computations, as most of the QSOs investigated here are chosen under similar conditions. This factor will therefore be included in the optical depth, deduced in Section 4.

## 3. Angular Selection Function

Several important issues must be taken into consideration when we define the angular selection function (ASF) (Surdej et al. 1993). First we must estimate the limitations on resolution and dynamical range (detectable magnitude differences), affecting our ability to resolve close components and large magnitude differences, respectively. The selection function will then simply describe our observational window in the angular separation-magnitude difference plane. We deduce the ASF for our particular survey in the following subsections.

### 3.1. Small separations

Fig. 2 illustrates the distribution of the full width at half maximum (FWHM) values for each of the 229 observations of the 168 objects. The seeing is distributed in a range between $0.5 - 2.0$ arc-second, indicating the highly variable image quality typical of ground-based observatories. This implies that each potential GL candidate has a different detectability range in the angular separation-magnitude difference plane. These differences obviously affect the possibility of detecting small separation GL-candidates ($\theta < 1''.0$). It is therefore necessary to weigh each observed image by its image quality (seeing-FWHM value). This is done by estimating the maximum detectable magnitude difference between two point-like components as a

---

[1] In fact bearing in mind the highly variable seeing and poor resolution of sky-plates, from which the QSOs are often chosen, the incompleteness accusations of QSO catalogues seem sensible.

**Table 1.** The complete observational list of the 168 QSOs in the NOT sample. See the notes at the end of this table for a description of the columns.

| Object | RA$_{1950}$ | Dec$_{1950}$ | F | FWHM | z | $m_V$ | $\Delta m$ | Sep | GLc | Note |
|---|---|---|---|---|---|---|---|---|---|---|
| H0000-263 | 000049.45 | -261959.2 | I | 1.11 | 4.11 | 17.5 | | | 0 | 1 |
| H0000+158 | 000325.0 | 155304 | R | 1.5 | 0.45 | 16.4 | | | 0 | 1 |
| H0007+1041 | 000728.06 | 104125 | R | 0.71 | 2.2 | 18.0* | | | 0 | 2,3 |
| H0017+154 | 001750.5 | 152423 | R | 1.3 | 2.01 | 18.2 | | | 0 | 1 |
| H0029+073 | 002943.35 | 072200.0 | R | 0.71 | 3.26 | 18.4 | | | 0 | 1 |
| H0033+156 | 003318.2 | 153655 | R | 1.5 | 1.16 | 18.0 | | | 0 | 1 |
| H0035+0725 | 003545.99 | 072532.6 | R | 0.58 | 3.0 | 17.8 | | | 0 | 3 |
| H0037+1206 | 003724.43 | 120654.2 | R | 0.5 | 2.5 | 18.5* | | | 0 | 3 |
| H0055-2659 | 005532.5 | -265925 | R | 1.1 | 3.67 | 17.5 | | | 2 | 2 |
| H0105-2634 | 010548.19 | -263420.2 | R | 1.45 | 3.5 | 17.5 | | | 2 | 2 |
| H0131+154 | 013129.7 | 152744 | R | 0.84 | 1.33 | 17.3 | | | 0 | 2 |
| H0143-0050 | 014339 | -005038 | R | 0.74 | 3.1 | 17.0 | | | 0 | 1 |
| H0146+0146 | 014644 | 014228 | R | 0.93 | 2.9 | 17.0 | | | 0 | 1 |
| HS0202+1848 | 020241.9 | 184811 | R | 1.2 | 2.71 | 17.5 | | | 0 | 1 |
| HS0211+18 | 021143.2 | 185840 | R | 1.2 | 2.48 | 18.0 | | | 0 | 1 |
| H0211+1118 | 021159.13 | 111804.0 | R | 1.07 | 2.0 | 16.0 | | | 0 | 2,3 |
| H0214+0820 | 021420.81 | 082007.1 | R | 1.16 | 2.0 | 18.0 | | | 0 | 2 |
| TEX0215+165 | 021555.3 | 163223.8 | V,I,R | 1.13 | 1.9 | 18.0 | | | 0 | 1 |
| H0216+0803 | 021618.1 | 080341 | R | 0.66 | 2.993 | 18.1 | | | 0 | 2 |
| H0218+0931 | 021813.52 | 093153.5 | R | 1.13 | 2.0 | 16.0 | | | 0 | 2,3 |
| HS0227+0558 | 022742.0 | 055829 | R | 0.71 | 2.06 | 17.5* | | | 0 | 1 |
| Q0229+1309 | 022902.63 | 130941 | V,R,I | 1.2 | 2.07 | 17.7 | 4.0 | 6.7 | 0 | 1 |
| Q0308+1902 | 030851.8 | 190225 | V,I,R | 0.86 | 2.84 | 18.6 | 3.1 | 4.4 | 1 | 1 |
| H0428-1342 | 042820 | -134212 | R | 1.26 | 3.2 | 17.0 | | | 0 | 1 |
| H0432-1246 | 043207.2 | -124608 | R | 1.05 | 2.8 | 17.0 | | | 0 | 1 |
| H0443-1702 | 044340.6 | -170243.4 | R | 0.9 | 1.6 | 17.0 | | | 0 | 1 |
| H0446-1440 | 044619 | -144027 | R | 1.42 | 1.6 | 17.0 | | | 0 | 1 |
| H0449-1325 | 044924 | -132532 | R | 1.0 | 3.1 | 17.0 | | | 0 | 1 |
| H0449-1645 | 044959 | -164509 | R | 1.30 | 2.6 | 17.0 | | | 0 | 2 |
| H0450-1310 | 045054 | -131039 | R | 1.07 | 2.3 | 16.5 | | | 0 | 1 |
| HS0621+6738 | 062138.5 | 673837 | R | 0.78 | 1.57 | 18.0* | | | 0 | 1 |
| HS0624+6907 | 062435.2 | 690703 | R | 0.83 | 0.37 | 13.9 | 5.8 | 8.0 | 0 | 1 |
| HS0626+6745 | 062655.3 | 674552 | R | 0.86 | 0.23 | 18.0* | 2.5 | 10.0 | 0 | 1 |
| Q0636+6801 | 063647.6 | 680127 | R | 0.94 | 3.18 | 16.5 | | | 0 | 1 |
| Q0640.3+4489 | 064018.2 | 445339 | V,R | 0.9 | 1.805 | 18.2 | -3.0 | 6.6 | 0 | 1 |
| B20650+37 | 065035.2 | 370927.1 | R | 0.9 | 1.982 | 18.0 | | | 0 | 1 |
| HS0701+6405 | 070130.3 | 640545 | R | 0.92 | 1.92 | 18.0 | | | 0 | 1 |
| HS0710+6024 | 071023.4 | 602441 | R | 0.94 | 1.77 | 18.0 | | | 0 | 1 |
| HS0727+6342 | 072746.2 | 634218 | R,I | 0.88 | 2.38 | 18.5 | 3.8 | 4.7 | 1 | 1 |
| HS0734+5956 | 073403.0 | 595643 | R | 0.94 | 1.75 | 16.5 | | | 0 | 1 |
| HS0734+6226 | 073413.0 | 622658 | R | 0.92 | 1.08 | 18.0 | | | 0 | 2 |
| H0738+3119 | 073800.2 | 311902.1 | R | 1.14 | 0.63 | 16.1 | 4.4 | 6.5 | 0 | 1 |
| HS0740+3222 | 074022.1 | 322219 | R | 1.1 | 1.52 | 18.0* | | | 0 | 1 |
| HS0742+3150 | 074230.7 | 315016 | V,R | 1.04 | 0.46 | 16.0 | | | 0 | 1 |
| HS0742+3320 | 074246.2 | 332055 | R | 0.94 | 0.61 | 17.7 | 2.5 | 9.9 | 0 | 1 |
| HS0743+6601 | 074358.6 | 660100 | R | 0.9 | 2.19 | 17.0 | | | 0 | 1 |
| HS0748+3321 | 074841.0 | 332104.0 | R | 1.12 | 1.932 | 18.5 | | | 0 | 1 |
| B20751+29 | 075151.0 | 294952 | R | 1.04 | 2.106 | 18.5 | | | 0 | 1 |
| HS0751+6107 | 075158.9 | 610747 | V,R,I | 1.3 | 2.61 | 17.5 | 3.75 | 5.6 | 0 | 1 |
| TEX0759+341 | 075933.4 | 340746 | R | 0.9 | 2.44 | 18.5 | | | 0 | 1 |
| HS0804+6218 | 080401.7 | 621826 | R | 0.9 | 1.13 | 17.5 | | | 0 | 1 |
| OJ508 | 080458.3 | 495923 | R | 0.9 | 1.433 | 17.5 | | | 0 | 1 |
| B20808+28 | 080832.2 | 285402 | R | 0.9 | 1.91 | 18.0 | | | 0 | 1 |

| Object | RA$_{1950}$ | Dec$_{1950}$ | F | FWHM | z | $m_V$ | $\Delta m$ | Sep | GLc | Note |
|---|---|---|---|---|---|---|---|---|---|---|
| H0820+1209 | 082037 | 120928 | R | 0.85 | 1.7 | 18.0* | | | 0 | 1 |
| H0821+1022 | 082127 | 102245 | R | 1.02 | 2.185 | 18.5* | | | 0 | 1 |
| H0822+0849 | 082217 | 084941 | R | 0.85 | 1.8 | 17.5* | | | 0 | 2 |
| H0828+1229 | 082830.3 | 122955 | R | 0.8 | 2.8 | 18.0* | | | 0 | 1 |
| H0831+1248 | 083123.11 | 124852.5 | R | 0.94 | 2.8 | 17.8 | | | 0 | 2 |
| H0834+1056 | 083405.4 | 105645 | R | 0.74 | 1.9 | 18.0* | | | 0 | 1 |
| H0834+0937 | 083434.4 | 093746 | R | 0.8 | 2.24 | 18.0* | | | 0 | 1 |
| H0836+1122 | 083649.0 | 112242 | R | 0.86 | 2.67 | 18.8 | | | 0 | 1 |
| CSO199 | 083807.6 | 355542 | R | 1.16 | 1.775 | 16.0 | | | 0 | 1 |
| HS0839+2858 | 083929.3 | 285811 | R | 1.27 | 1.34 | 18.0 | | | 0 | 1 |
| H0841+1256 | 084138.9 | 125643 | R | 0.73 | 2.2 | 18.0 | | | 0 | 1 |
| CSO203 | 084230.5 | 343154 | R | 0.96 | 2.126 | 17.0 | | | 0 | 1 |
| H0843+1454 | 084317.17 | 145423.9 | R | 0.86 | 2.28 | 18.0 | | | 0 | 1 |
| HS0843+2743 | 084333.1 | 274344 | R | 1.4 | 2.03 | 18.5 | | | 0 | 1 |
| H0845+1523 | 084549.76 | 152310.8 | R | 0.72 | 2.4 | 18.0 | | | 0 | 1 |
| HS0845+3013 | 084559.9 | 301344 | R | 0.92 | 0.67 | 17.0* | | | 0 | 2 |
| H0846+1540 | 084620.58 | 154039.5 | R | 0.93 | 2.9 | 18.0 | | | 0 | 2,3 |
| H0846+1517 | 084650.95 | 151733.2 | R | 0.86 | 2.6 | 18.0 | | | 0 | 1 |
| RQ0849+2845 | 084905.8 | 284516 | R | 0.92 | 1.27 | 20.2 | | | 0 | 1 |
| RQ0849+2820 | 084948.7 | 282001 | R | 0.94 | 0.2 | 18.5 | | | 0 | 2 |
| HS0850+2852 | 085022.3 | 285233 | V,I,R | 1.2 | 2.09 | 17.5 | 4.5 | 3.7 | 1 | 1 |
| H0850+1755 | 085047.2 | 175512.2 | R | 0.72 | 3.21 | 18.0 | | | 0 | 1 |
| H0853+1953 | 085335.45 | 195312.0 | R | 1.23 | 2.8 | 18.0 | | | 0 | 1 |
| H0854+1632 | 085459.37 | 163236.6 | R | 0.88 | 2.54 | 18.0 | | | 0 | 1 |
| HS0855+2549 | 085503.1 | 254928 | R | 0.9 | 1.80 | 18.0 | 4.3 | 3.3 | 1 | 1 |
| H0903+1534 | 090305.39 | 153450.3 | R | 0.92 | 2.8 | 18.0 | | | 0 | 1 |
| H0904+2003 | 090429.86 | 200316.9 | R | 1.0 | 2.54 | 18.5 | | | 0 | 1 |
| H0905+1507 | 090537.49 | 150724.7 | R | 0.92 | 3.16 | 19.0 | 2.7 | 3.1 | 1 | 1 |
| CSO233 | 093632.3 | 365348 | R | 1.06 | 2.03 | 17.0 | | | 0 | 2 |
| HS0945+4630 | 094536.1 | 463029 | R | 1.0 | 0.99 | 17.5* | | | 0 | 2 |
| HS0945+4646 | 094555.0 | 464614 | R | 1.2 | 1.91 | 18.0* | | | 0 | 1 |
| HS0948+4631 | 094814.6 | 463154 | R | 1.11 | 1.8 | 18.0* | | | 0 | 1 |
| HS1003+4733 | 100325.0 | 473345 | R | 1.15 | 0.27 | 17.5* | | | 0 | 2,3 |
| HS1004+4515 | 100431.4 | 451531 | R | 1.2 | 1.3 | 17.0* | | | 0 | 1 |
| HS1004+4543 | 100449.4 | 454306 | R | 1.15 | 1.66 | 17.5* | | | 0 | 1 |
| CSO38 | 100904.6 | 295648 | R | 1.0 | 2.62 | 16.0 | | | 0 | 2,3 |
| CSO261 | 101653.6 | 355654 | R | 1.04 | 2.67 | 17.0 | | | 0 | 1 |
| Q1107+487 | 110748.2 | 484730 | R | 0.77 | 3.00 | 16.7 | | | 0 | 1 |
| CSO340 | 112332.6 | 353636 | R | 0.93 | 1.285 | 17.0 | | | 0 | 1 |
| B31123+395 | 112345.8 | 393515.1 | V,I,R | 1.06 | 1.47 | 16.4 | 2.5 | 5.4 | 0 | 1 |
| Q1123+434 | 112349.4 | 432607.4 | R | 1.06 | 2.01 | 18.4 | | | 0 | 1 |
| CBS42 | 112839.8 | 292054 | R | 1.11 | 1.319 | 17.0 | | | 0 | 1 |
| US2694 | 113417.2 | 300808 | V,I,R | 1.14 | 1.858 | 18.3 | | | 2 | 2 |
| PG1148+549 | 114842.6 | 545413 | R | 0.85 | 0.969 | 16.1 | | | 0 | 2 |
| PC1158+4635 | 115802.9 | 463529 | R | 1.2 | 4.733 | 22.4 | 1.5 | 8.9 | 0 | 1 |
| Q12076+399 | 120738.6 | 395617 | R | 1.3 | 2.4 | 17.5 | 4.0 | 7.3 | 0 | 2,3 |
| Q1210+1731 | 121030.4 | 173104 | R | 0.87 | 2.537 | 17.4 | | | 0 | 3 |
| Q1222.7+22.6 | 122247.2 | 223704 | V,I,R | 1.15 | 2.29 | 18.0 | 0.8 | 4.3 | 1 | 1 |
| Q1223+1753 | 122335.8 | 175325 | R | 0.92 | 2.918 | 18.1 | | | 0 | 1 |
| Q1230+1627 | 123039.4 | 162726 | R | 0.98 | 2.7 | 17.8 | | | 0 | 1 |
| CSO151 | 123126.9 | 292424 | R | 1.2 | 2.011 | 16.0 | | | 0 | 2 |
| SP1 | 135806.0 | 390854 | R | 0.75 | 3.28 | 17.0 | | | 0 | 2 |
| UM627 | 135837.0 | 000124 | R | 0.98 | 1.87 | 16.0 | | | 0 | 2 |
| HS1404+32 | 140445 | 320331 | R | 1.35 | 2.37 | 17.5* | | | 0 | 2 |
| CSO609 | 140635.7 | 491624 | R | 0.95 | 2.13 | 17.0 | | | 0 | 2,3 |
| Q1412+0925 | 141225.6 | 092510 | R,I | 1.05 | 1.70 | 17.4 | 2.5 | 6.3 | 0 | 2 |

| Object | RA$_{1950}$ | Dec$_{1950}$ | F | FWHM | z | $m_V$ | $\Delta m$ | Sep | GLc | Note |
|---|---|---|---|---|---|---|---|---|---|---|
| GB21413+373 | 141322.7 | 372015 | R | 1.54 | 2.36 | 18.0 | | | 0 | 1 |
| HS1420+326 | 142021.1 | 323647 | R | 0.94 | 0.69 | 17.5 | | | 0 | 2,3 |
| HS1421+330 | 142117.5 | 330555 | R | 1.03 | 1.9 | 16.5 | | | 0 | 2,3 |
| TEX1421+359 | 142144.1 | 355606 | R | 0.98 | 1.57 | 17.5 | | | 0 | 2 |
| Q1429+1153 | 142947.2 | 115306 | R | 1.85 | 3.01 | 18.6 | | | 0 | 1 |
| Q1435-0134 | 143513.2 | -013413 | R | 1.7 | 1.31 | 16.0 | | | 0 | 1 |
| Q1442+295 | 144244.4 | 293142 | V,R,I | 1.7 | 2.67 | 16.2 | 2.0 | 3.4 | 1 | 1 |
| CSO722 | 150445.0 | 542324 | V,I,R | 0.84 | 1.9 | 17.0 | 2.4 | 6.4 | 0 | 1 |
| LB9612 | 151708.2 | 235652 | R | 0.96 | 1.9 | 17.3 | | | 0 | 1 |
| SP43 | 152030.0 | 412212 | R | 0.84 | 3.1 | 17.5 | | | 0 | 1 |
| LB9707 | 152308.8 | 212436 | R | 1.1 | 1.924 | 18.0 | | | 0 | 1 |
| TEX1558+187 | 155802.9 | 184654 | R | 0.9 | 2.4 | 18.0 | | | 0 | 1 |
| PKS1559+17317 | 155904.6 | 172236 | R | 1.05 | 1.944 | 18.0 | | | 0 | 1 |
| B31621+392 | 162123.4 | 391627 | V,I,R | 0.6 | 1.97 | 17.5 | 2.3 | 3.8 | 1 | 1 |
| Q1623+268 | 162345.4 | 265354 | R | 1.15 | 2.52 | 16.0 | | | 0 | 1 |
| HS1626+64 | 162620.4 | 643332 | R | 0.92 | 2.32 | 16.5* | 3.2 | 9.0 | 0 | 1 |
| Q1628+3808 | 162828.0 | 380447 | R | 0.73 | 1.461 | 16.8 | | | 0 | 1 |
| Q1630+3749 | 163032.2 | 373925 | R | 0.73 | 2.037 | 18.2 | | | 0 | 1 |
| Q1633.5+26.7 | 163334.8 | 264417 | R,I | 1.16 | 1.84 | 17.0 | | | 0 | 1 |
| PG 1634+706 | 163451.7 | 703737 | R | 2.24 | 1.33 | 14.9 | | | 0 | 2 |
| HS1638+61 | 163846.8 | 612153 | R | 0.88 | 0.46 | 18.0* | | | 0 | 1 |
| Q1643+401 | 164326.7 | 400443 | R | 0.84 | 1.877 | 18.0 | | | 0 | 1 |
| HS1700+6416 | 170040.5 | 641625 | R | 0.8 | 2.74 | 16.1 | | | 2 | 2 |
| HS1704+6048 | 170403.5 | 604831 | R | 1.1 | 0.37 | 15.3 | | | 0 | 1 |
| 1E1704+710 | 170500.6 | 710134.0 | R | 1.0 | 2.00 | 17.5 | | | 0 | 1 |
| HS1709+6027 | 170951.4 | 602723 | R | 1.1 | 1.54 | 17.0* | | | 0 | 1 |
| PG1715+535 | 171530.7 | 533124 | V,I,R,Z,B | 0.72 | 1.93 | 16.5 | 0.116 | 3.7 | 1 | 1 |
| HS1716+6839 | 171627.9 | 683948 | R | 1.7 | 0.78 | 18.5 | | | 0 | 1 |
| PG1718+481 | 171817.7 | 480711 | R | 1.45 | 1.08 | 14.7 | | | 0 | 1 |
| Q1720+3470 | 172002.3 | 344149 | V,I,R | 1.5 | 2.4 | 18.4 | 1.3 | 7.7 | 0 | 1 |
| Q1722+34 | 172259.8 | 344437 | R | 1.2 | 2.2 | 18.5 | 1.9 | 6.1 | 0 | 1 |
| HS1723+6550 | 172307.2 | 655026 | R | 1.15 | 1.45 | 17.0* | | | 0 | 1 |
| TEX1726+344 | 172601.3 | 342504 | I | 1.6 | 2.425 | 18.5 | | | 0 | 1 |
| HS1732+6535 | 173246.2 | 653523.0 | R | 0.97 | 0.86 | 17.6 | | | 0 | 1 |
| HS1742+6147 | 174221.6 | 614711.0 | R | 1.22 | 0.52 | 18.6 | | | 0 | 1 |
| HS1749+7006 | 174903.4 | 700639 | R | 0.94 | 0.77 | 17.4 | | | 0 | 1 |
| HS1803+6737 | 180337.4 | 673754 | R | 1.9 | 0.14 | 15.8 | | | 0 | 1 |
| HS1807+6948 | 180718.5 | 694857.1 | R | 1.8 | 0.051 | 14.4 | | | 0 | 1 |
| PKS1821+10 | 182141.6 | 104244 | R | 1.03 | 1.36 | 17.3 | 3.4 | 8.0 | 0 | 1 |
| HS1824+6507 | 182435.9 | 650737 | V,B,R,I | 1.75 | 0.3 | 16.5* | 1.5 | 5.2 | 0 | 1 |
| 4C56.28 | 185730.1 | 564146 | V,Z,R,I | 1.2 | 1.595 | 17.3 | 2.7 | 3.2 | 1 | 1 |
| HS1946+7658 | 194641.0 | 765826 | V,I,R | 1.2 | 3.02 | 15.9 | | | 0 | 1 |
| PC2047+0123 | 204750.7 | 012356 | R | 1.10 | 3.799 | 19.7 | | | 0 | 1 |
| TEX2048+196 | 204856.7 | 193849 | R | 0.85 | 2.36 | 18.5 | | | 0 | 1 |
| TEX2116+203 | 211617.2 | 202122 | V,R,I | 0.98 | 1.68 | 17.0 | 2.75 | 3.7 | 1 | 1 |
| TEX2127+176 | 212717.7 | 174136 | V,R,Z,I | 0.9 | 2.01 | 18.0 | 1.3 | 2.1 | 1 | 1 |
| TEX2127+348 | 212746.2 | 345324.0 | R,I | 1.6 | 2.4 | 18.5 | 2.7 | 8.1 | 0 | 1 |
| PKS2150+050 | 215054.1 | 052209 | R | 0.9 | 1.979 | 17.8 | | | 0 | 1 |
| H2230+114 | 223007.81 | 112822.5 | R | 0.75 | 1.04 | 16.5 | | | 0 | 1 |
| Q2231+0125 | 223142.1 | 012541 | R | 0.9 | 1.9 | 18.2 | | | 0 | 1 |
| H2251+158 | 225129.4 | 155254.0 | R,I | 0.7 | 0.86 | 15.3 | | | 0 | 1 |
| H2251+113 | 225140.7 | 112036.8 | R | 1.51 | 0.32 | 15.8 | | | 0 | 1 |
| TEX2300+345 | 230059.7 | 343038 | R | 0.7 | 2.490 | 18.0 | | | 0 | 1 |
| B22310+383 | 231036.3 | 383123.0 | R | 1.6 | 2.17 | 17.5 | | | 0 | 1 |
| H2310+0018 | 231050.8 | 001826.4 | R | 0.83 | 2.2 | 17.0* | | | 0 | 2,3 |

| Object | RA$_{1950}$ | Dec$_{1950}$ | F | FWHM | z | $m_V$ | $\Delta$m | Sep | GLc | Note |
|---|---|---|---|---|---|---|---|---|---|---|
| H2314+160 | 231409.7 | 160143.5 | R | 1.5 | 0.66 | 18.0 | | | 0 | 2 |
| H2355+0845 | 235521.6 | 084521.6 | R | 0.7 | 2.8 | 18.0* | | | 2 | 2 |
| Q2359+0653 | 235906.77 | 065313.3 | R | 0.45 | 3.24 | 18.8 | | | 0 | 2,3 |
| H2359+1305 | 235957.4 | 130532 | R | 0.77 | 2.8 | 18.0* | | | 0 | 3 |

*Notes*:
*Col. 1*: QSO identification, where the id indicates list of origin; H - from Hazard, HS - from Reimers (Hamburg Sternwarte) and others from the Véron catalogue
*Col. 2*: QSO RA (hhmmss) in equatorial 1950 coord.
*Col. 3*: QSO Dec (ddmmss) in equatorial 1950 coord.
*Col. 4*: Filter types
*Col. 5*: Seeing (FWHM) in arcsec
*Col. 6*: Redshift z of the QSO
*Col. 7*: V-band magnitude of the QSO as estimated from our data or tabulated in other QSO catalogues. The QSOs with no previously measured magnitude were estimated by comparing them to QSOs on other CCD frames (for a similar exposure time). These estimated magnitudes are marked by an asterix (∗) and have errors up to ±0.5 mag
*Col. 8*: Magnitude difference between closest companion with weaker apparent brightness and within 5″ of QSO
*Col. 9*: Separation (in arc second) of closest companion with weaker apparent brightness and within 5″ of QSO
*Col.10*: A "1" indicates that the QSO is treated as a possible GL candidate, while a "2" indicates that the object is elongated, diffuse or otherwise especially interesting. The latter objects are not treated as GL candidates in the present analysis, but should be further analysed in future work
*Col. 11*: A "3" indicates that there are no PSFs other than the QSO in the image, a "2" indicates that the QSO appears elongated (this is sometimes due to bad tracking) and a "1" indicates that there are one or more PSFs and the QSO is not elongated function of their angular separation, which we evaluate by performing a simulation-test. Using a typical point spread function (PSF) for different seeing values, we have produced simulated images of various closely separated pair-scenarios, adopting a random flux ratio. The results of these tests are illustrated in Fig. 3, which also show the selection function for some of the other surveys being added in Section 6. We discuss these tests in more detail in Section 3.3.

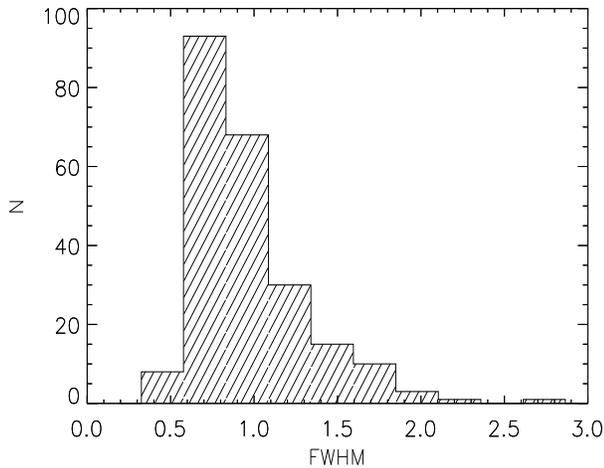

**Fig. 2.** The seeing (FWHM in arc-sec.) distribution of 229 observations of 168 QSOs

### 3.2. Large separations

In the identification of possible GL-candidates we chose all QSOs with a companion weaker than the QSO within the area $\pi r_c^2$. If more than one close companion exists, we chose the brighter one, because it has the highest probability of being a gravitationally lensed image of the QSO. We have adopted a cutoff radius $r_c = 5$ arc-second to secure the inclusion of most GL-systems produced by a single isolated galaxy and exclude those produced by clusters of galaxies. Using a larger cutoff radius would also include many star or galaxy companions and, hence, increase the number of false positives (Kochanek 1993b).

### 3.3. The ASF

To determine the ASF of the NOT sample for a wide range of seeing values we performed detection tests on synthetic data. These data were produced from numerical PSFs obtained from one or several stars in an image of a certain quality (we used PSFs with FWHM of $0''.6$, $0''.75$, $1''.0$ and $1''.2$). The synthetic images consisted of a closely separated image-pair of PSFs, which were produced using random angular-separation and magnitude difference. Photon (Poisson) noise and read-out noise were included in these simulations. Inspection by eye was performed for a series of synthetic images in each FWHM category. For each of these categories, a set of constants ($C_1$, $C_2$, $A$ and $B$) were determined, describing the particular ASF for that particular seeing (FWHM) value. Cubic interpolation was applied to determine these constants for the intermediate seeing values, thereby estimating the ASF for any given seeing value within reasonable range of the actual measured values. The ASF is valid for seeing values from these tests in the range $0\rlap{.}''5 \leq$ FWHM $\leq 1\rlap{.}''5$. We also found from these tests that the detection probabilities are slightly dependent on the position angle (PA) of the image pair for small separations. Specifically, alignments along the rows or columns of the detector decrease the probability of separating the pair on the basis of inspection by eye.

The complete ASF, which we apply in our work, is now represented by

$$\Delta m(\theta) = \begin{cases} \text{no solution} & \text{for } 0'' < \theta \leq C_1 \\ B + A \times \theta & \text{for } C_1 < \theta \leq C_2 \\ 5.0 \text{ mag} & \text{for } C_2 < \theta \leq r_c \end{cases} \quad (1)$$

where $C_1$ and $C_2$ are determined for each FWHM value and constitute the starting value and "break" point of the ASF, respectively. Some of the constants that determine the ASF for various values of seeing (FWHM) are listed in Table 2. The ASF is also shown in Fig. 3.

**Table 2.** Constants describing the ASF for various seeing (FWHM) values.

| FWHM | A | B | $C_1$ | $C_2$ |
|---|---|---|---|---|
| $0\rlap{.}''5$ | 22.318 | -4.305 | 0.192 | 0.416 |
| $0\rlap{.}''6$ | 17.700 | -3.520 | 0.198 | 0.482 |
| $0\rlap{.}''7$ | 14.193 | -3.185 | 0.224 | 0.576 |
| $0\rlap{.}''8$ | 12.845 | -3.536 | 0.276 | 0.664 |
| $0\rlap{.}''9$ | 13.390 | -4.508 | 0.336 | 0.710 |
| $1\rlap{.}''0$ | 14.176 | -5.712 | 0.402 | 0.756 |
| $1\rlap{.}''1$ | 13.809 | -6.854 | 0.496 | 0.858 |
| $1\rlap{.}''2$ | 12.599 | -8.000 | 0.636 | 1.032 |
| $1\rlap{.}''3$ | 11.526 | -9.385 | 0.814 | 1.248 |

## 4. Statistical model

Following previous GL statistical studies such as Turner et al. (1984),hereafter TOG;Fukugita & Turner (1991),hereafter F&T;Surdej et al. (1993); Kochanek (1993c) and Maoz & Rix (1993), we have used the well known singular isothermal sphere (SIS) model in the statistical analysis of the combined SCYM and NOT samples. In this section we outline the SIS model as well as the magnification bias.

A SIS galaxy model is characterised by its constant deflection angle (independent of the impact parameter) giving the critical (Einstein) radius $b_{cr} = 4\pi\sigma^2/c^2(D_{ls}/D_s)$, where $D_{ls}$ and $D_s$ are the proper distances[2] lens-source and observer-source, respectively. A "strong lensing" event occurs when the QSO is projected within the area defined by the critical radius, $\pi b_{cr}^2$. This area denotes the cross section for lensing by a SIS lens. Because the one dimensional central velocity dispersion, $\sigma$, is the only free parameter in this model, the statistical analysis is very sensitive to $\sigma$. Recently, Kochanek (1993c) proposed that the correction factor of $\sqrt{3/2}$ to the $\sigma$ value, which is commonly applied in lens statistics to account for the presence of DM in late-type galaxies (TOG, F&T), should not be used. We will discuss this in Section 6 and present results for both cases (with and without the DM correction).

---
[2] The critical radius $b_{cr}$ can be expressed equally using either angular *or* proper distances (Kochanek 1993a).

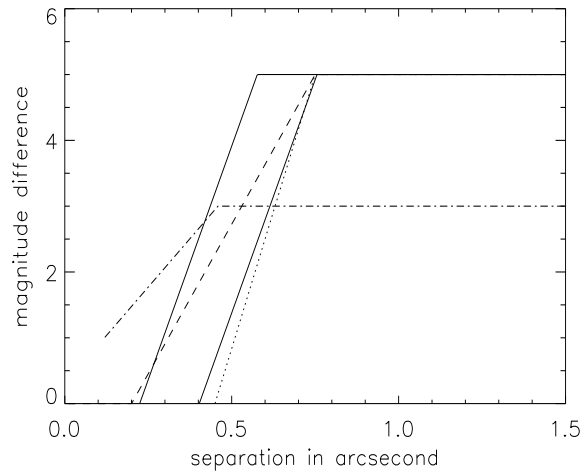

**Fig. 3.** The ASF for NOT observations at $0\rlap{.}''7$ and $1\rlap{.}''0$ seeing (FWHM) (solid lines). Also included, for comparison, are the ESO-KP ASFs for ground-based observations with inspection by EYE (dotted) and PSF-subtraction (dashed) for 1.0" seeing and the ASF for Hubble Space Telescope (HST) snapshot survey (dash-dot) (see Surdej et al. 1993).

As usual, we estimate the population of SIS galaxies from the Schechter galaxy luminosity function

$$\phi_g dL = n_* \left(\frac{L}{L_*}\right)^\alpha \exp(-\frac{L}{L_*}) dL/L_*, \quad (2)$$

where the galaxy number density $n_* = 1.56(\pm 0.4) \times 10^{-2} h^3$ Mpc$^{-3}$ and $\alpha = -1.1 \pm 0.1$, following Efstathiou et al. (1988). The Hubble constant is $H_0 = 100 h$ km s$^{-1}$ Mpc$^{-1}$. The galaxy luminosity, $L$, is estimated by using the Faber-Jackson relation

$$L = L_* \left(\frac{\sigma}{\sigma_*}\right)^\gamma \Rightarrow \sigma = \sigma_* \left(\frac{L}{L_*}\right)^{1/\gamma}, \quad (3)$$

where the exponent is taken to be $\gamma = 4$ (Tully & Fisher 1977; Faber & Jackson 1976). The value of the exponent $\gamma$ is chosen (rather than using the tabulated values of 3.7, 3.7 and 2.6 for E, S0 and S galaxies, respectively) for computational reasons and do not affect significantly the statistical results (Kochanek 1993c). Furthermore, we assume that the number of lenses is constant in a co-moving (i.e. non-evolving) volume, hence the density $n_l$ of lenses is given by $n_l = n_0(1 + z_l)^3$, where the $(1 + z_l)^3$ factor arises from the fact that the co-moving volume decreases with increasing $z$, due to the expansion of the universe. The present density of isolated galaxies, $n_0$, is found from integrating the Schecter luminosity function Eq. (2) over all luminosities (i.e. $n_0 = \int_0^\infty \phi_g dL$).

**Table 3.** The relative abundance, central velocity dispersions and computed F values of late-type E/S0 galaxies and early-type S spiral galaxies, using $\alpha = -1.1$ and $\gamma = 4$ (Kochanek 1993c). The values of and uncertainties in the abundance, $n_*$ and $\sigma_*$ are taken from F&T (and references therein). Note that the F values are computed using $n_* = 1.56(\pm 0.4) \times 10^{-2}\ h^3$ Mpc$^{-3}$ and with/without the DM correction factor of $\sqrt{3/2}$.

| type | abund. | $\sigma_*\ [kms^{-1}]$ | F | $\sigma_*\ [kms^{-1}]$ | F |
|---|---|---|---|---|---|
| E | 12% | $225^{+12}_{-20}$ | $0.0085 \pm 0.0044$ | $225^{+12}_{-20} \times \sqrt{3/2}$ | $0.0191 \pm 0.0099$ |
| S0 | 19% | $206^{+12}_{-20}$ | $0.0094 \pm 0.0058$ | $206^{+12}_{-20} \times \sqrt{3/2}$ | $0.0212 \pm 0.0115$ |
| S | 69% | $143^{+8}_{-13}$ | $0.0080 \pm 0.0042$ | $143^{+8}_{-13}$ | $0.0080 \pm 0.0042$ |
| E/S0/S | 100% | | $0.0259 \pm 0.0084$ | | $0.0483 \pm 0.0157$ |
| E/S0 | 31% | | $0.0179 \pm 0.0073$ | | $0.0403 \pm 0.0151$ |

The optical depth, $\tau$, for lensing of a distant QSO with magnitude $m_V$, and redshift $z_s$, is given by

$$\tau = \frac{1}{30} F D_s^3(z_s) \tag{4}$$

where $F = \frac{16\pi^3}{cH_0^3}\langle n_0\sigma^4\rangle = \frac{16\pi^3}{cH_0^3}\Gamma[1+\alpha+4/\gamma]n_*\sigma_*^4$ and $\Gamma$ is the gamma-function (TOG, F&T).

In Table 3 we list all relevant abundances, velocity dispersions and respective F parameter values with/without DM corrections for both E/S0 and E/S0/S populations. Note that the DM corrections are only applied to late-type galaxies (i.e. E/S0).

In our statistical model we must also take into consideration the important magnification bias, which increases the expected number of lenses. This is due to the magnification of (strongly) lensed images, which effectively alters the magnitude limited sample (Narayan & Wallington 1993). We estimate the magnification bias directly from the QSO luminosity function, which we define as in F&T (based on data from Hartwick & Schade (1990))

$$n_q(m_B) = \begin{cases} 10^{0.86(m_B-19.15)} & \text{for } m_B < 19.15 \\ 10^{0.28(m_B-19.15)} & \text{for } m_B \geq 19.15 \end{cases} \tag{5}$$

Here $n_q(m_B)$ represents the counts of QSOs at apparent magnitude $m_B$. The differential magnification bias, simply being the over-representation of QSOs at magnitude $m_B$ due to the magnification of originally dimmer QSOs with magnitudes $m_B + 2.5\log M$, is given by

$$B(m_B, M) = \frac{dP_M}{dM} \frac{n_q(m_B + 2.5\log M)}{n_q(m_B)} \tag{6}$$

where $dP_M/dM = 8/M^3$ is the magnification (M) probability distribution for our particular SIS lens model, assuming the QSO is a point source.

## 5. Results from the NOT sample

We here present the objectively discovered, possible candidates of gravitational lensing from the 168 QSO NOT sample. The candidates within the boundaries of the NOT ASF are listed in Table 4. Each candidate is also discussed individually and additional colour and spectral data, which might enable us to reject or approve a candidate, are also presented here. Some objects have been observed by other surveys, in which case we present their conclusions. Since only a few candidates have been observed spectroscopically at NOT using the low dispersion spectrograph (LDS), not many of the candidates have been rejected. Furthermore, because no photometric-calibration exposures were taken for the NOT sample, we are not able to supply colour estimates. We have, however, compared the magnitude difference between the QSO and nearby companions for the various filters and use similar magnitude differences as an indication of colour resemblance.

- *PG 1715+535*'s companion is confirmed by Yee et al. (1993) and Crampton et al. (1992) to be a star, having also been investigated spectroscopically by these surveys. Its measured colour differences ($\Delta m_V = 0.12$, $\Delta m_R = 0.12$ and $\Delta m_I = -0.05$) could not reject either interpretation, which is why it could only be determined spectroscopically. The small magnitude difference (i.e. high magnification) is the main contribution to the high lensing probability.
- *TEX 2127+176* and its companion have a significant colour difference ($\Delta m_V = 1.3$, $\Delta m_R = 0.5$ and $\Delta m_I = 0.3$) and is therefore not a strong candidate. Spectroscopic observations were unfortunately not properly achieved. Because of the small angle separating the QSO ($m_V \sim 18$) and companion ($m_V \sim 19.5$), we suspect that the companion was not satisfactorily aligned on the slit. We were able to identify the rather strong CIV emission line in the spectra of the QSO, while nothing was detectable at that location in the fainter spectrum. Further spectroscopic observations are planned in 1995 and will hopefully confirm or reject the lensing hypothesis.
- *Q 1442+295* and its companion show a significant colour difference ($\Delta m_V = 2.0$, $\Delta m_R = 1.6$ and $\Delta m_I = 1.1$), indicating that the companion is quite redder than the QSO. Spectroscopic data show no sign of a distinct alikeness in emission line or continuum. We conclude that this object is not a strong GL candidate, most probably a star. Recent spectroscopic observations of this pair, carried out at the

**Table 4.** List of possible GL-candidates in the NOT-sample, with redshift, V band magnitude, angular separation and magnitude difference. Objects are sorted in decreasing relative calculated probability of the companion being a lensed image of the QSO.

| Object: | $z_s$ | $m_V$ | $\theta$ | $\Delta$ m | comment |
|---|---|---|---|---|---|
| PG1715+535 | 1.93 | 16.50 | 3″.7 | 0.12 | star |
| TEX2127+176 | 2.01 | 18.00 | 2″.1 | 1.30 | red color |
| Q1442+295 | 2.67 | 16.20 | 3″.4 | 2.00 | red color |
| TEX2116+203 | 1.68 | 17.00 | 3″.7 | 2.75 | red color |
| Q1222.7+22.6 | 2.29 | 18.00 | 4″.3 | 0.80 | spectra needed |
| 4C 56.28 | 1.59 | 17.30 | 3″.2 | 2.70 | red color |
| H0905+1507 | 3.16 | 19.00 | 3″.1 | 2.70 | spectra needed |
| B3 1621+392 | 1.97 | 17.50 | 3″.8 | 2.30 | star |
| Q0308+1902 | 2.84 | 18.60 | 4″.4 | 3.10 | galaxy |
| HS0855+2549 | 1.80 | 18.00 | 3″.3 | 4.30 | |
| HS0850+2852 | 2.09 | 17.50 | 3″.7 | 4.50 | |
| HS0727+634 | 2.38 | 18.50 | 4″.7 | 3.80 | spectra needed |

<u>Notes</u>:
<i>Col. 1</i>: HLQ identification
<i>Col. 2</i>: Redshift z of the QSO
<i>Col. 3</i>: V-magnitude of the QSO as tabulated in other QSO catalogues
<i>Col. 4</i>: Angular separation $\theta$ between the QSO and its companion (if more than one, chosen as described in Section 2)
<i>Col. 5</i>: Magnitude difference in the V-band between the QSO and companion
<i>Col. 6</i>: Comment on components nature or characteristics

NTT, La Silla, Chile by J. Surdej also show little resemblance between the QSO and companion (Surdej 1994).

– *TEX 2116+203* and its companion show a significant colour difference ($\Delta m_V = 2.75$, $\Delta m_R = 2.2$ and $\Delta m_I = 1.5$). Spectroscopic observations give no indication that the companion is a lensed image of the QSO.
– *Q 1222.7+22.6* has only been imaged in the R and V filters. The magnitude differences in these colours are, however, quite similar ($\Delta m_V \approx 0.8$ and $\Delta m_R \approx 0.65$) and within what we expect to be plausible. The QSO remains, therefore, an interesting candidate of lensing.
– *4C 56.28* and its companion have been observed by other surveys, but have not yet been investigated spectroscopically. Our photometric data indicate that the companion is somewhat redder than the QSO (in agreement with previous surveys), but not more than can be expected from a reddening effect due to a lensing galaxy ($\Delta m_V \approx 2.7$ and $\Delta m_R \approx 2.65$). We therefore treat this pair as interesting and plan to perform spectroscopic observations in 1995.
– *H 0905+1507*'s companion was observed spectroscopically by J.Surdej at the NTT and seems to be quite red, though no spectroscopic data is yet available (Surdej 1994).
– *B3 1621+392* and its companion show fairly similar colours ($\Delta m_V \approx 2.3$, $\Delta m_R \approx 2.25$ and $\Delta m_I \approx 2.4$), but the companion has been reported by Bahcall et al. (1992) to be a G-type star.
– *Q 0308+19* and its companion show fairly similar colours ($\Delta m_V \approx 3.1$, $\Delta m_R \approx 3.0$ and $\Delta m_I \approx 3.0$), but the companion has been reported by Crampton et al. (1992) to be a galaxy.
– *HS 0855+25* lacks photometric data and demands further investigation if its companion's nature is to be revealed.
– *HS 0850+28* and its companion show fairly similar colours ($\Delta m_V \approx 4.5$, $\Delta m_R \approx 4.5$ and $\Delta m_I \approx 4.5$), but has not yet been observed spectroscopically.
– *HS 0727+63* and its companion show fairly similar colours ($\Delta m_R \approx 3.8$ and $\Delta m_I \approx 3.5$), but has not yet been observed spectroscopically.

## 6. Cosmological implications

To make the best possible statistical analysis (i.e. using as large a sample as possible) we combine the NOT sample with the full optical SCYM-sample presented partially [3] by Surdej et al. (1993). Using likelihood functions, we estimate maximum likelihood velocity dispersions, $\sigma_*$, and lens effectiveness parameters, $F$, from the combined sample of 784 HLQs for various flat cosmologies. In these computations we employ complete integral probabilities. This is the probability for a given QSO (with apparent magnitude, $m_i$, and redshift, $z_i$) to be lensed by an intervening isolated galaxy.

We now wish to express the probability of finding a lensed image of a QSO under the constraints imposed by the observational conditions. The observational conditions are characterised by the seeing (FWHM) of point sources in the images. The detectable magnitude difference is dependent on the image angular separation, and is closely related to the seeing. We apply the ASF introduced in Section 3, which describes the relations of detectable magnitude differences, angular separation and seeing for our particular survey. Following Surdej et al. (1993) and Kochanek (1993c), we can express the probability, $p_i^{SF}$, of finding a gravitationally lensed image in the vicinity (within $r_c$) of a QSO within the observational window (defined by the selection function) by

$$p_i^{SF}(m_i, z_i) = \frac{S_{cat}F}{30} D_s^3(z_i)$$
$$\int_0^{r_c/2} p_c(b_{cr}) Bias(m_i, M(2b_{cr}), M_2) db_{cr}. \quad (7)$$

Here the critical radius probability is given by $p_c(b_{cr}) = \frac{d\tau}{\tau db_{cr}} = \frac{2d\tau}{\tau d\theta}$ (refer Eq. (4.10) in Kochanek (1993c)) [4]. $Bias(m_i, M, M_2)$ is the integral of $B(m_i, M)$, given by Eq. (6), and the bias integral, using an infinite upper limit, is numerically computed by substitution of variables [5].

---
[3] We have also included, as did Kochanek (1993c), the extended HST snapshot sample presented in Maoz et al. (1993).
[4] We have expressed the lensing probability in terms of the critical radius, $b_{cr}$, instead of the angular separation $\theta$.
[5] There was little or no difference in our results using either this method or that of setting the upper limit to a large number, i.e. $M_2 = 10000$

**Table 5.** Estimated values of $\sigma_*$ and F from the frequency of lensing and the configuration likelihoods of the observations. The results are shown in each row for different combinations of galaxy population with and without DM correction. The 90% confidence level results are shown in the upper part of the table for the Einstein-de Sitter universe model ($\Omega_M = 1$). In the lower part, we give the 90% confidence level results for the maximum likelihood value of the density parameter (i.e. $\Omega_M = 0.95$ for E/S0/S and $\Omega_M = 0.25$ for E/S0 populations). The maximum likelihood value is indicated by $F^{max}$ and $\sigma_*^{max}$, respectively.

| $\Omega_M$ | galaxy types | $F^{low}$ | $F^{max}$ | $F^{high}$ | $\sigma_*^{low}$ | $\sigma_*^{max}$ | $\sigma_*^{up}$ |
|---|---|---|---|---|---|---|---|
| 1.00 | E/S0/S | 0.02021 | 0.03735 | 0.06589 | 164.5 | 191.8 | 221.1 |
| 1.00 | E/S0/S (without DM) | 0.00898 | 0.01660 | 0.02928 | 134.3 | 156.6 | 180.5 |
| 1.00 | E/S0 | 0.01490 | 0.02477 | 0.04031 | 204.3 | 232.0 | 262.0 |
| 1.00 | E/S0 (without DM) | 0.00662 | 0.01101 | 0.01792 | 166.8 | 189.4 | 213.9 |
| 0.95 | E/S0/S | 0.01964 | 0.03615 | 0.06360 | 163.3 | 190.3 | 219.1 |
| 0.95 | E/S0/S (without DM) | 0.00873 | 0.01607 | 0.02826 | 133.4 | 155.3 | 178.9 |
| 0.25 | E/S0 | 0.00609 | 0.01111 | 0.01944 | 163.3 | 189.9 | 218.3 |
| 0.25 | E/S0 (without DM) | 0.00271 | 0.00494 | 0.00864 | 133.4 | 155.0 | 178.3 |

Note also that $Bias(m_i, M(2b_{cr}), M_2)$ is normalised so that $\int_{min(M)}^{M_2} dP_M/dM\, dM = 1$. The minimum detectable magnification (i.e. maximum magnitude difference) is given by the expression for the magnification $M(2b_{cr}) = 2\frac{f(2b_{cr})+1}{f(2b_{cr})-1}$, where $f(2b_{cr}) = f(\theta)$ is the flux ratio corresponding to the detectable magnitude difference $\Delta m(\theta)$ (given by Eq. (1)). In our calculations, we have taken advantage of the "Numerical Recipes in Fortran" implementation of the Romberg method for integration (Press et al. 1992). Note also that in absence of any selection function the minimum detectable magnification would be $2.5 \log 2$ (TOG, Surdej et al. 1993).

The probability for finding a lensed pair with a certain redshift and magnitude difference within the ASF, is thus the probability for lensing weighted by the ASF (Eq. (1)) for each critical radius $b_{cr}$. The configuration probability $p_c(b_{cr})$ (the relative probability of a separation) multiplied by the ratio of the bias including the ASF to that with no ASF, is thus integrated over the survey area ($r \leq r_c$) and multiplied by the general probability of lensing, leading to Eq. (7). The sum of the probabilities of finding a GL-system within the detectable ranges of the selection function for each observed object, $\sum_i p_i^{SF}$, will then give us the expected number of lensed QSOs in the sample.

The combined NOT+SCYM sample, consisting of 784 QSOs, has been applied in the computations below. Among these QSOs, four have been identified as being gravitationally lensed: the two double imaged lensed systems "Q 1208+1011" (Magain et al. 1988; Maoz et al. 1992) and "UM 673" (Surdej et al. 1987), the clover-leaf "H1413+117" (Magain et al. 1988) and the triple image "PG 1115+080" (Weymann et al. 1980), which were all found in the SCYM-sample. We therefore have $N_L = 4$. The well known "QSO 0957+561" (Walsh et al. 1979) is not within our cutoff radius and is therefore not among the GL-candidates for the survey. It is also a well known fact that the gravitational potential of this GL-system is not solely due to the massive central galaxy, but also due to the surrounding cluster (contributing to the large separation of 6.1 arc-second). The *four* lensed systems are included, because even though our lens model is not able to reproduce the image configuration of two of them, the frequency of lensing is still preserved.

Kochanek (1993c) estimated $\sigma_*$ for E and S0 galaxy populations without the DM correction factor, while TOG and F&T computed the lens effectiveness parameter, F with the DM correction factor. Surdej et al. (1993) estimated the F parameter without á priori corrections for DM. Since we do not know the importance of S spiral galaxies nor that of dark matter, we estimate $\sigma_*$ for both populations and with/without the DM correction factor of $\sqrt{3/2}$. Furthermore, we wish to investigate if it is possible to obtain reasonable constraints on the cosmological density parameter, $\Omega_M$.

In these computations we apply the configuration probability likelihood function introduced in Kochanek (1993c)

$$P_{tot} = \prod_{i=1}^{N_U}(1 - p_i^{SF}) \prod_{j=1}^{N_L} p_j^{SF} \prod_{k=1}^{N_L} p_{ck} \qquad (8)$$

which, converted into logarithmic likelihoods, becomes

$$ln P_{tot} = -\sum_{i=1}^{N_U} p_i^{SF} + \sum_{j=1}^{N_L} ln p_j^{SF} + \sum_{k=1}^{N_L} ln p_{ck} \qquad (9)$$

under the assumption that $p^{SF} \ll 1$. It is here assumed that $p_i$, $p_j$ and $p_{ck}$ are the probabilities of a QSO being not lensed, lensed or having a separation $b$ if lensed, respectively. The configuration probability of a certain separation $b$ is given by

$$p_{ck} = p_c(b_{cr}) \frac{\tau_*(\sigma_*) D_s^3(z_s)}{p_k^{SF}} Bias(m_k, M(b_{cr}), M_2), \qquad (10)$$

while $p_i^{SF}$ and $p_j^{SF}$ are given by Eq. (7). The logarithmic likelihood function adds together all probabilities and counts lensed HLQs as positive and unlensed HLQs as negative contributions.

We, hence, compute likelihoods for various values of the velocity dispersion, $\sigma_*$, and flat cosmological models (i.e. different $\Omega_M = 1 - \Omega_\Lambda$ values). The value of $\Omega_M$ affects the proper distance, which is part of the optical depth (Eq. (4)),

A.O. Jaunsen et al.: The NOT GL survey for multiply imaged quasars    11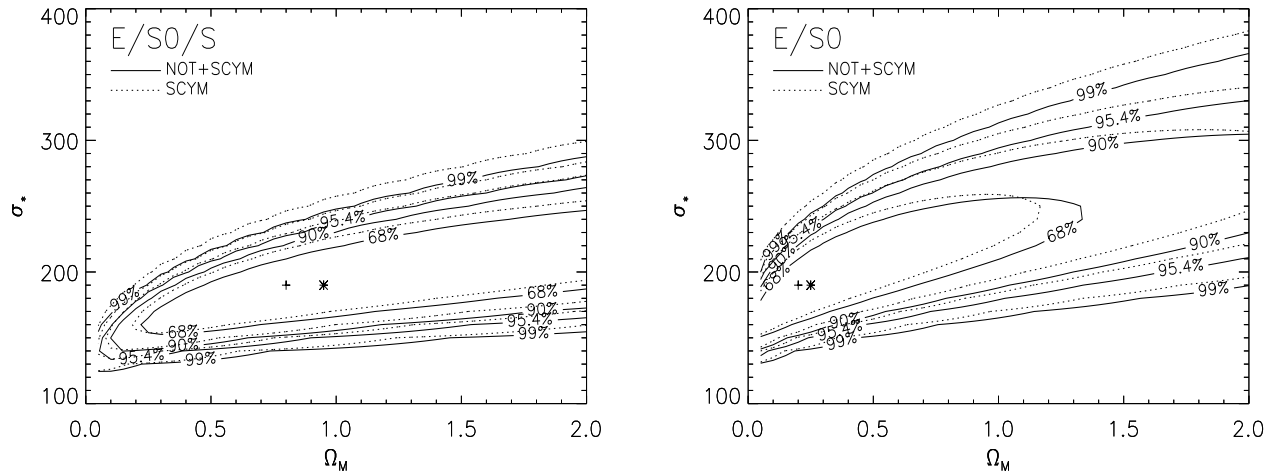

**Fig. 4.** Likelihood contour plots for various flat cosmologies ($\Omega_M = 1 - \Omega_\Lambda$) and velocity dispersions ($\sigma_*$), assuming lensing is significant for all lens types (E/S0/S) (left) and only for late type galaxies (E/S0) (right). Both the NOT+SCYM and SCYM samples are represented in the plots by solid and dotted lines, respectively. The maximum likelihood model for these two samples are indicated by ∗ and +, respectively.

while $\sigma_*$ governs the $F$ parameter (the optical depth $\tau$) and the image separation distribution, $p_c(b_{cr})$. The individual lensing probability $p_i^{SF}$ is computed for each object for 1600 various combinations of $\Omega_M$ and $\sigma_*$ (a 40 × 40 array).

We present the likelihood results from these computations as contours of the 68%, 90%, 95% and 99% confidence levels in Fig. 4. These two-dimensional confidence levels are simply the corresponding inverse percentage levels of the maximum likelihood. The computation of confidence intervals for the $\sigma_*$ or $F$ parameter (a single parameter), can be determined from the $\chi^2$ distribution with one degree of freedom. The 90% confidence intervals are therefore determined from the 26% peak values of the two dimensional likelihood contours in Fig. 4 (see §5 in Kochanek (1993c) or Lupton (1993), p.70).

In Fig. 4 we have plotted the various confidence levels for the NOT+SCYM sample as well as those for just the SCYM sample. Note that all of the identified lenses were discovered in the SCYM sample, so there is no loss in the number of lenses when reducing the sample from NOT+SCYM to SCYM. There is, however, a slight difference in the results, specifically (and not surprisingly) the uncertainties are somewhat lower for the NOT+SCYM sample. Also, because there is a smaller number of lenses pr. QSO in the NOT+SCYM sample, the contours show that the typical $\sigma_*$ values are somewhat lower than for the SCYM sample.

We present, in Table 5, the confidence intervals for the F parameter and $\sigma_*$ in the case of lensing by all galaxy types (E/S0/S) and for late-type galaxies (E/S0). Our results are in best agreement with the models (in Table 3) excluding the DM correction factor. The results are, however, too uncertain to draw any definite conclusions about the DM correction factor. If, however, the DM correction factor is excluded, the effect of S galaxies is not negligible and should be included as part of the lensing population (see Section 4).

If we inspect the diagrams of Fig. 4, we realize that trying to constrain the cosmological models ($\Omega_M$), will still give too uncertain conclusions. The 90% confidence levels on $\sigma_*$ (without DM) from our results, however, give $164.5 < \sigma_* < 221.1$ kms$^{-1}$ (for $\Omega_M = 1.0$) and $204.3 < \sigma_* < 262.0$ kms$^{-1}$ (for $\Omega_M = 1.0$) for E/S0/S and E/S0 galaxy populations, respectively.

To investigate roughly how the uncertainties are decreasing with an increased number of QSOs, we have made simulations using synthetic samples by simply increasing the present number of QSOs and lenses by a factor, $n$, of two, five and ten, albeit this obviously introduces some unwanted selection effects. These tests show that the uncertainties on $\sigma_*$ diminish approximately proportionally to $\sqrt{n}$, while the uncertainties in $\Omega_M$ diminishes somewhat more rapidly.

Conclusively we find it difficult to discriminate between various flat cosmologies based on the present statistical model and sample, while the possible values of galaxy velocity dispersions, $\sigma_*$, are more restricted. Further examinations of enlarged samples of HLQs and QSOs in addition to improved statistical analysis will definitely better discriminate between the various parameters.

*Acknowledgements.* The authors would like to thank Jean Surdej for critically commenting and thereby greatly improving on the manuscript. A.O. Jaunsen would also like to thank J. Surdej, S.V.H.Haugan, C. Kochanek, A.A. Kaas, S. Wallington and J.F. Claeskens for valuable help during this work. This work was partially supported by NFR, the Norwegian Research Council and by EC HCM Network CHRX-CT92-0044.